\documentclass[%
 reprint,
superscriptaddress,
 amsmath,amssymb,
 aps,
prb,
floatfix,
]{revtex4-2}

\usepackage{graphicx}
\usepackage{dcolumn}
\usepackage{bm}

\usepackage[colorlinks,
           urlcolor=blue,
           linkcolor=blue,
           anchorcolor=blue,
           citecolor=blue
           ]{hyperref}
\usepackage{float}
\usepackage{xcolor}
\usepackage{soul}
\usepackage{siunitx}

\begin{document}

\preprint{APS/123-QED}

\title{Predictions and Measurements of Thermal Conductivity of Ceramic Materials at High Temperature}


\author{Zherui Han}
 \altaffiliation{These authors contributed equally to this work.}
 \affiliation{School of Mechanical Engineering and the Birck Nanotechnology Center,\\
Purdue University, West Lafayette, Indiana 47907-2088, USA}

\author{Zixin Xiong}
 \altaffiliation{These authors contributed equally to this work.}
 \affiliation{School of Mechanical Engineering and the Birck Nanotechnology Center,\\
Purdue University, West Lafayette, Indiana 47907-2088, USA}

\author{William T. Riffe}
\affiliation{Department of Materials Science and Engineering, University of Virginia, Charlottesville, Virginia 22904, USA}

\author{Hunter B. Schonfeld}
\affiliation{Department of Mechanical and Aerospace Engineering, University of Virginia, Charlottesville, Virginia 22904, USA}
 
\author{Mauricio Segovia}
\affiliation{School of Mechanical Engineering and the Birck Nanotechnology Center,\\
Purdue University, West Lafayette, Indiana 47907-2088, USA}

\author{Jiawei Song}
\affiliation{School of Materials Engineering, Purdue University, West Lafayette, Indiana 47907-2088, USA}
 
\author{Haiyan Wang}
 \affiliation{School of Materials Engineering,
Purdue University, West Lafayette, Indiana 47907-2088, USA}

 \author{Xianfan Xu}
 \affiliation{School of Mechanical Engineering and the Birck Nanotechnology Center,\\
Purdue University, West Lafayette, Indiana 47907-2088, USA}

\author{Patrick E. Hopkins}
 \affiliation{Department of Mechanical and Aerospace Engineering, University of Virginia, Charlottesville, Virginia 22904, USA}
 \affiliation{Department of Materials Science and Engineering, University of Virginia, Charlottesville, Virginia 22904, USA}
 \affiliation{Department of Physics, University of Virginia, Charlottesville, Virginia 22904, USA}

\author{Amy Marconnet}
 \email{marconnet@purdue.edu}
 \affiliation{School of Mechanical Engineering and the Birck Nanotechnology Center,\\
Purdue University, West Lafayette, Indiana 47907-2088, USA}

\author{Xiulin Ruan}%
 \email{ruan@purdue.edu}
 \affiliation{School of Mechanical Engineering and the Birck Nanotechnology Center,\\
Purdue University, West Lafayette, Indiana 47907-2088, USA}

\date{\today}

\begin{abstract}
The lattice thermal conductivity ($\kappa$) of two ceramic materials, cerium dioxide (CeO$_2$) and magnesium oxide (MgO), is computed up to 1500~K using first principles and the phonon Boltzmann Transport Equation (PBTE) and compared to time-domain thermoreflectance (TDTR) measurements up to 800~K. Phonon renormalization and the four-phonon effect, along with high temperature thermal expansion, are integrated in our \textit{ab initio} molecular dynamics (AIMD) calculations. This is done by first relaxing structures and then fitting to a set of effective force constants employed in a temperature-dependent effective potential (TDEP) method. Both three-phonon and four-phonon scattering rates are computed based on these effective force constants. Our calculated thermal conductivities from the PBTE solver agree well with literature and our TDTR measurements. Other predicted thermal properties including thermal expansion, frequency shift, and phonon linewidth also compare well with available experimental data. Our results show that high temperature softens phonon frequency and reduces four-phonon scattering strength in both ceramics. Compared to MgO, we find that CeO$_2$ has weaker four-phonon effect and renormalization greatly reduces its four-phonon scattering rates.

\end{abstract}


\maketitle


\section{Introduction}

High temperature properties are important for many energy and power technologies primarily for two reasons. First, increasing operating temperature can enhance the power generation or conversion efficiency~\cite{Perepezko2009HotterBetter}, and potentially lower the cost of renewable energy~\cite{Henry2014CSP}. Second, power generations involving thermochemical reactions usually require sufficiently high temperature to sustain appreciable power density~\cite{Shin2020MaterialsHighT}. Thermal transport in materials at high temperature ($\sim1000$~K or above) plays an indispensable role in these processes~\cite{Shin2020MaterialsHighT}. In particular, thermal barrier coatings (TBCs)~\cite{Padture2002TBC} protect structural components enabling them to survive the high operating temperature. Candidate materials for TBCs are usually ceramics as they have low thermal conductivity, high thermal stability, and appropriate thermal expansion behavior~\cite{Klemens1998ThermalTBC,Shin2020MaterialsHighT}.

In dielectric solids, heat is mainly carried by phonons, the quantum mechanical description of lattice vibrations, and their dynamics can be described by the phonon Boltzmann Transport Equation (PBTE)~\cite{peierls1929kinetischen,ziman1960electronsphonons}. The thermal conductivity of ceramics is then limited by various interaction processes including intrinsic three-phonon scattering and scattering by defects and boundaries as discussed previously in the work of Klemens and Gell~\cite{Klemens1998ThermalTBC}. In the recent decade, the advent of first-principles technique, coupled with the PBTE, gives accurate computation of phonon-phonon scattering rates and lattice thermal conductivity~\cite{broido2007intrinsic}. This approach based on the lowest-order perturbation, three-phonon scattering, has achieved great success in predicting thermal properties~\cite{Esfarjani2011Si,seko2015prediction,lindsay2013BAs,review2018mtp,JAPreviewMcGaughey}. However, theoretical models that work well at moderate temperatures face new challenges at high temperatures. Measurements at high temperatures on many technologically important materials like silicon and diamond often report thermal conductivities lower than predicted~\cite{4ph2017}. On the other hand, another study reports nearly 50\% underprediction in uranium oxide at 1500~K~\cite{Pang2013UO2Phonon}. Measurements of phonon linewidth, a direct observation of optical phonon anharmonicity, show the opposite trend with temperature compared to predictions~\cite{prl2007anharm,prb2010lifetimeExp,prb2011measurement}. In light of these comparisons, higher-order anharmonicity is expected to have significant impact on thermal transport~\cite{4ph2016,4ph2017,highthroughput2020prx}. This four-phonon effect is expected to be stronger at higher temperatures as it could scale quadratically with temperature~\cite{4ph2017}. Furthermore, phonon renormalization or the finite-temperature effect is necessary to correct the potential landscape and renormalize the phonon energies when the traditional quasiharmonic approximation (QHA) fails~\cite{TDEP2011,CSLD}. High temperature transport is likely to significantly deviate from QHA and needs to be considered from this new perspective. Several theoretical studies on highly anharmonic crystals have been devoted to integrating both four-phonon scattering and the finite-temperature effect~\cite{xia2018revisiting,ravichandran2018unified}. These emerging theories motivate us to investigate high temperature thermal transport in ceramics and extend the earlier theoretical understanding~\cite{Klemens1998ThermalTBC}.

In this work, we combine \textit{ab initio} molecular dynamics (AIMD) with the PBTE to evaluate thermal properties of two ceramics, cerium dioxide (CeO$_2$) and magnesium oxide (MgO), at temperatures up to 1500~K. We compare our predicted thermal properties to available high temperature experiments and our own time-domain thermoreflectance (TDTR) measurements from room temperature to 800~K. Investigated thermal properties include thermal expansion, optical phonon frequencies, phonon scattering rates (or linewidths), and thermal conductivity. Such detailed and comprehensive comparisons with experiments that are rarely seen in previous theoretical reports enable us to examine and verify different levels of physics. Note that radiation component of thermal transport is not discussed here as we focus on lattice thermal conduction~\cite{Klemens1998ThermalTBC}.

Through our first-principles calculations, we find that both CeO$_2$ and MgO have positive lattice thermal expansion. As temperature increases, the potential landscape is greatly affected: phonon frequencies are softened and phonon scattering rates are weakened. Comparison between two ceramics shows that at high temperatures four-phonon effect is not strong for CeO$_2$ but well exceeds three-phonon scattering in MgO.

\section{Methodology}

\subsection{Effect of temperature}
We approach high temperature thermal transport in solids with a comprehensive view. Solution of the PBTE gives us thermal conductivity $\kappa$ in terms of phonon mode-wise contribution:
\begin{equation}
    \kappa = \sum_\lambda C_{\lambda}v_\lambda^2 \tau_\lambda,
\end{equation}
where $C_{\lambda}$, $v_\lambda$, and $\tau_\lambda$ are volumetric heat capacity, group velocity, and relaxation time for a phonon mode $\lambda$ having momentum $\mathbf{q}$. A temperature-dependent treatment without any prior assumptions requires that all elements in the calculation should be functions of temperature. In the above description, heat capacity is expressed as $C_{\lambda}=\frac{1}{V}(\partial (\hbar \omega n^0_\lambda)/\partial T) =\frac{k_{\rm B}}{V}(\hbar \omega/k_{\rm B} T)^2 n^0_\lambda(n^0_\lambda+1)$, which is a function of lattice volume $V$, temperature $T$, and phonon frequency $\omega$. The phonon population at equilibrium obeys the Bose-Einstein distribution: $n^0_\lambda=1/(e^{(\hbar \omega/k_{\rm B} T)}-1)$, where $k_{\rm B}$ is the Boltzmann constant. The group velocity $v_\lambda$ is determined by the phonon dispersion. The relaxation time $\tau_\lambda$ is the inverse of scattering rate and can be computed by Fermi's golden rule. We write the scattering rate of a three-phonon absorption process $\Gamma^{(+)}_{\lambda\lambda'\lambda''}$ as an example where one phonon $\lambda$ absorbs another phonon $\lambda'$ and emits a third phonon $\lambda''$ ($\lambda+\lambda'\to \lambda''$):

\begin{equation}
    \Gamma^{(+)}_{\lambda\lambda'\lambda''}=\frac{\hbar\pi}{4}\frac{n^0_{\lambda'}-n^0_{\lambda''}}{\omega_\lambda\omega_{\lambda'}\omega_{\lambda''}}\vert V_{\lambda\lambda'\lambda''}^{(+)}\vert^2\delta(\omega_\lambda+\omega_{\lambda'} -\omega_{\lambda''})\Delta^{(+)}.
\end{equation}
Conservation of energy is enforced by the Dirac delta function $\delta$ and the matrix elements $V_{\lambda\lambda'\lambda''}^{(+)}$ are given by the Fourier transformation of third-order interatomic force constants $\Phi_3$. The Kronecker delta $\Delta^{(+)}=\Delta_{\mathbf{q}+\mathbf{q'}+\mathbf{Q},\mathbf{q''}}$ describes the momentum conservation where $\mathbf{Q}$ is a reciprocal lattice vector with $\mathbf{Q}=0$ implying normal process. Similar expressions can be derived for four-phonon scattering with the computation of the fourth-order interatomic force constants $\Phi_4$. The calculation of phonon scattering rates gives us temperature dependence of $\tau_\lambda$ on phonon dispersion, phonon population, and force constants. Since phonon dispersion is solved by dynamical matrix encoded in the harmonic force constants $\Phi_2$, it is reasonable to argue that temperature dependence of a thermal transport property $\rho(T)$ is some function ($F$) of lattice expansion, phonon population, and different orders of force constants in the current computational formalism or
\begin{equation}
    \rho(T) = F(n(T), V(T), \Phi_2(T), \Phi_3(T),\Phi_4(T)).
    \label{rho-F}
\end{equation}

Conventional theory with three-phonon scattering and ground-state first-principles calculations is essentially a simplified version of Eq.~\ref{rho-F}:
\begin{equation}
    \begin{split}
    \rho(T) &= F_{\rm conv}(n(T),V(0~\textrm{K}),  \Phi_2(0~\textrm{K}), \Phi_3(0~\textrm{K})).
\end{split}
\end{equation}
This description gives a temperature dependence that is solely determined by change of occupation number. A more reasonable description given by Eq.~\ref{rho-F} is that elevated temperature expands or shrinks the lattice, renormalizes the quasiparticle phonon energy and changes the potential landscape in the system.

For the comparisons and discussions in later sections, we have calculated the set of force constants using 0~K potential landscape, i.e., $\Phi_2$, $\Phi_3$ and $\Phi_4$.

\subsection{\textit{Ab initio} molecular dynamics}

As the temperatures that we define as ``high temperatures'' usually exceed the Debye temperature of ceramic materials, molecular dynamics simulation is suitable to describe the lattice dynamics. Simulations are carried out in Vienna \textit{Ab initio} Simulation Package (\textsc{VASP})~\cite{VASP1993}. The first step of our calculation is to relax the structure at a certain temperature and get a new potential energy surface. Anharmonic force constants can then be evaluated on this new baseline. Under a constant-temperature, constant-pressure ensemble (NPT) with zero external pressure, the lattice relaxation is performed on a supercell structure constructed by $4\times4\times4$ primitive cells (192 atoms for CeO$_2$ and 128 atoms for MgO). On such a relaxed lattice, we switch to a constant-temperature, constant-volume ensemble (NVT). We employ a temperature-dependent effective potential (TDEP) method detailed in Ref.~\cite{TDEP2011,TDEP2013} that collects force-displacement dataset $\{\mathbf{F}_t^{\mathrm{MD}},\mathbf{U}_t^{\mathrm{MD}}\}$ in productive $N_t$ AIMD time steps to construct effective harmonic force constants $\Phi_2^{*}$. These effective harmonic force constants are obtained by minimizing force differences in MD and model harmonic system such that it can best describe potential surface of thermally excited lattice~\cite{TDEP2011,TDEP2013}:

\begin{equation}
\min _{\Phi_2^*} \Delta \mathbf{F}=\frac{1}{N_t} \sum_{t=1}^{N_t}\left|\mathbf{F}_t^{\mathrm{MD}}-\mathbf{F}_t^{\mathrm{H}}\right|^2,
\end{equation}
where $\mathbf{F}_t^{\mathrm{H}}$ is the force calculated from effective harmonic system $\mathbf{F}_t^{\mathrm{H}}=\Phi_2^*\mathbf{U}_t^{\mathrm{MD}}$ in the $t$-th time step. A similar minimization process can extend to higher-order effective force constants by fitting the residual term in harmonic system~\cite{TDEP2013IFCs} and we are able to obtain a set of effective higher-order force constants $\Phi_3^{*}$ and $\Phi_4^{*}$. In this work, the NVT simulation is performed 20000 time steps to reach equilibrium and the extraction of effective force constants is performed on a final productive run of 2000 time steps.

\subsection{Phonon scattering}

Given a set of force constants describing the potential energy surface, the phonon scattering rates are computed by Fermi's golden rule. In this study, we consider both three-phonon (3ph) and four-phonon (4ph) scattering. The full expression for scattering rate of one mode $\tau_\lambda^{-1}$ under relaxation time approximation (RTA) is a summation of all the channels based on Matthiessen's rule~\cite{ziman1960electronsphonons}:

\begin{widetext}
\begin{equation}
    \frac{1}{\tau_\lambda}=\frac{1}{N}\left(\displaystyle\sum^{(+)}_{\lambda'\lambda''}\Gamma^{(+)}_{\lambda\lambda'\lambda''}+\displaystyle\sum^{(-)}_{\lambda'\lambda''}\frac{1}{2}\Gamma^{(-)}_{\lambda\lambda'\lambda''}\right)+
  \frac{1}{N}\displaystyle\sum^{\text{(iso)}}_{\lambda'}\Gamma^{\text{(iso)}}_{\lambda\lambda'}+\frac{1}{N}\left(\displaystyle\sum^{(++)}_{\lambda'\lambda''\lambda'''}\frac{1}{2}\Gamma^{(++)}_{\lambda\lambda'\lambda''\lambda'''}+\displaystyle\sum^{(+-)}_{\lambda'\lambda''\lambda'''}\frac{1}{2}\Gamma^{(+-)}_{\lambda\lambda'\lambda''\lambda'''}+\displaystyle\sum^{(--)}_{\lambda'\lambda''\lambda'''}\frac{1}{6}\Gamma^{(--)}_{\lambda\lambda'\lambda''\lambda'''}\right),
  \label{eq.phononscattering}
\end{equation}
\end{widetext} 
where $N$ is the total number of gridpoints of $\mathbf{q}$ points when solving PBTE. The superscripts $(\pm)$ and $(\pm\pm)$ represent the 3ph and 4ph processes, repectively. The notations are for 3ph absorption ($\lambda+\lambda'\to \lambda''$) and emission ($\lambda \to \lambda'+ \lambda''$), and 4ph recombination ($\lambda+\lambda'+\lambda''\to \lambda'''$), redistribution ($\lambda+\lambda'\to \lambda''+\lambda'''$) and splitting ($\lambda\to\lambda'+\lambda''+\lambda'''$) processes. $\Gamma^{\text{(iso)}}$ is the isotope scattering rates. The calculation of Eq.~\ref{eq.phononscattering} requires $\Phi_2^{*}$, $\Phi_3^{*}$ and $\Phi_4^{*}$ obtained from AIMD simulations and is performed in a modified version our computational program, FourPhonon~\cite{han2021fourphonon}. In our calculation, primitive cell of CeO$_2$ is sampled by $10\times10\times10$ $q-$mesh and that of MgO is sampled by $12\times12\times12$ $q-$mesh.

\subsection{TDTR measurement}

We use time-domain thermoreflectance (TDTR) to characterize thermal conductivity of MgO and CeO$_2$. A laser beam with a repetition rate of 80~MHz from a Ti:Sapphire oscillator is split into pump and probe beams. The pulsed pump beam periodically heats the surface of the sample, which is coated with a thin ($\sim80$~nm) layer of Al transducer. A probe beam delayed up to 5.5~ns after heating by a pump pulse measures the change in reflectance due to change in the temperature detected at the surface. The pump beam is modulated by an electro-optic modulator (EOM) at 8.4~MHz. The focused pump and probe beams have $1/e^2$ diameters of $\sim19$ and $\sim10$~$\mu$m, respectively. Each sample is coated with a thin layer of aluminum (the transducer layer) which serves to deposit the heating at the surface and enable measurement of the thermal response through a known $dR/dT$ where $R$ is the surface reflectance. To obtain thermal conductivity and interface conductance, we fit the magnitude of the probe beam ($\sqrt{V_{in}^2+V_{out}^2}$) to a model of the response due to accumulative periodic heating by a pulsed laser~\cite{TTR,FeldmanTDE,CahillTDTR}. Heat capacities and densities of materials are taken from literature~\cite{MgOproperties,Nelson2014Ceo2stochiometric}. We fit the thermoreflectance response with two free parameters for the MgO substrate: the thermal conductivity of the substrate and the thermal conductance between the substrate and the transducer layer. For the CeO$_2$ thin film, there are three free parameters: the thermal conductivity of the film, the thermal conductance between the film and the transducer, and the thermal conductance between the film and the substrate. To ensure repeatability, we record data on different days at several locations for each sample. Above room temperature, samples are heated and stabilized in a thermal stage with flowing Argon gas to maintain a clean chamber environment. 

\section{Results}
Here, we present our first-principles results on various high temperature thermal properties. In each subsection we compare our results to available experimental reports or our own measurements.

\subsection{Thermal expansion}

\begin{figure*}[ht]
\centerline{\includegraphics[width=6in]{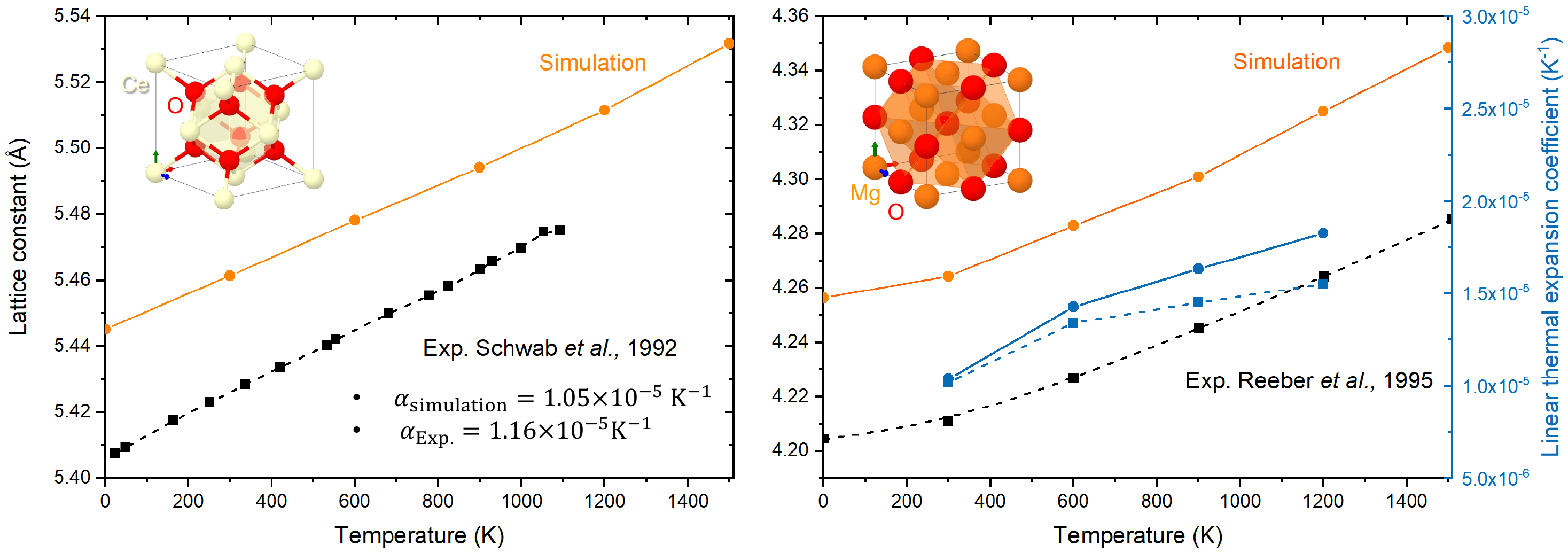}}
\caption{Lattice constants of (a) CeO$_2$ and (b) MgO as a function of temperature. Simulation results are presented in orange solid lines, while experimental data~\cite{Schwab1992CeriaExpTEC,Reeber1995MgOExpTEC} are in black with dashed lines to guide the eye. In the right panel, blue lines (simulation in solid and experiment in dashed line) correspond to the linear thermal expansion coefficient of MgO in right $y-$axis that varies with temperature. Atomic structures in this figure are generated from a toolkit of Materials Project~\cite{MaterialsProject,Toolkit}.}
\label{fig.thermalexpansion}
\end{figure*}

Figure~\ref{fig.thermalexpansion} shows our calculated lattice constants as a function of temperature for both ceramics. We find that the lattice constant increases nearly linearly with temperature in CeO$_2$ resulting in a nearly constant thermal expansion coefficient, while the thermal expansion coefficient of MgO varies with temperature. The Debye temperature of CeO$_2$ is lower than that of MgO, so that the quantum effect is more pronounced in MgO. The predicted behaviors are consistent with experimental reports~\cite{Schwab1992CeriaExpTEC,Reeber1995MgOExpTEC}.
For CeO$_2$, the simulated linear thermal expansion coefficient is $\alpha=1.05\times10^{-5}~\mathrm{K}^{-1}$ while the experimental reported value is around $1.16\times10^{-5}~\mathrm{K}^{-1}$~\cite{Schwab1992CeriaExpTEC}. As for MgO, the trend with temperature is similar to measured values as quantified by the linear thermal expansion coefficient shown in blue lines. But the absolute values of lattice constants are off by a small fraction for both materials and it is likely due to an error in ground state (0~K) structure determined by Density Functional Theory (DFT). We note that the relative errors of lattice constants are within 1.5\% for both materials and such small difference is acceptable in first-principles calculations. For evaluation of thermal properties, the agreement of thermal expansion behavior is more essential than absolute lattice parameters.

\subsection{Phonon dispersions}

\begin{figure*}[!htb]
\centerline{\includegraphics[width=6.8in]{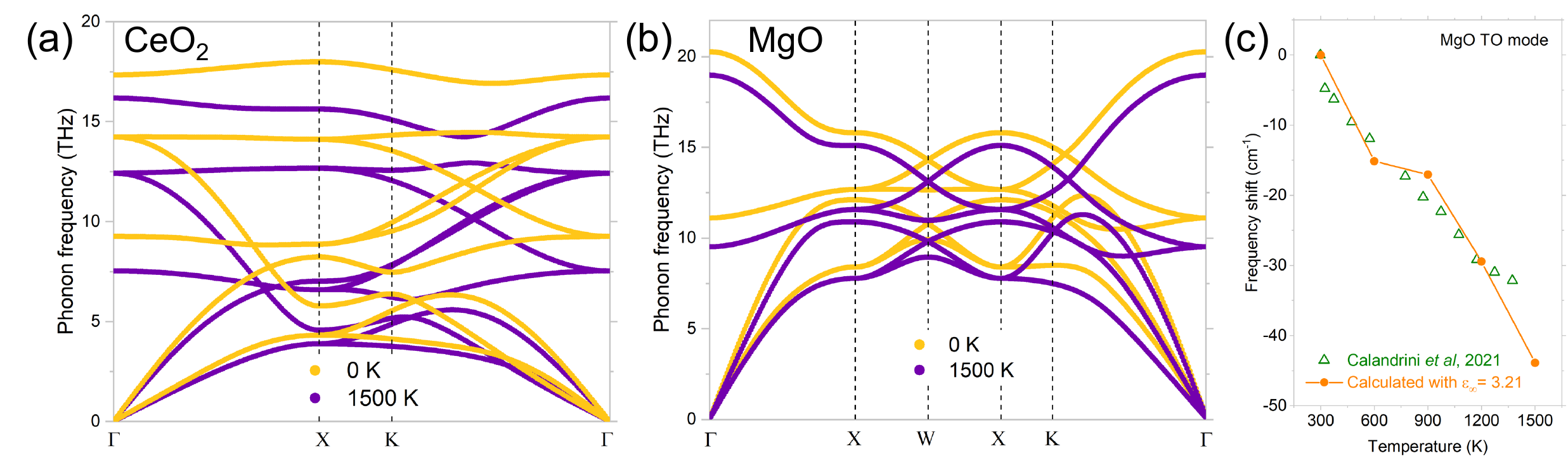}}
\caption{Phonon dispersions at 0~K and 1500~K for (a) CeO$_2$ and (b) MgO along high-symmetry paths. (c) Phonon frequency shift for TO phonon mode of MgO as a function of temperature. Experimental data from Ref.~\cite{Calandrini2021MgOOpticalModes} (green triangles) is compared to our calculated results (orange circles connected by orange lines). 
Note that our calculated high-frequency dielectric constant $\epsilon_{\infty}$ for MgO is 3.24.
}
\label{fig.phonon}
\end{figure*}

From the effective harmonic force constants $\Phi_2^{*}$ that are extracted from AIMD at each finite temperature, we can solve for phonon dispersions that are renormalized. Figure~\ref{fig.phonon} presents our calculated phonon dispersions at 1500~K compared to 0~K. Since CeO$_2$ and MgO are both polar materials, longer-range interactions are needed to capture the splitting of transversal and longitudinal optical phonons (TO and LO modes) so that they are not degenerate at the Brillouin zone-center. This interaction is computed by Density Functional Perturbation Theory (DFPT) and we get Born effective charge tensor $Z^*_{ij}$ and high-frequency dielectric constant $\epsilon_{\infty}$ for both materials.

Phonon energies are lower at higher temperature for all phonon modes in both materials. Our result is consistent with another theoretical study on CeO$_2$~\cite{Klarbring2018CeriaFiniteT}, but differs from a recent theoretical paper on MgO~\cite{Kwon2020MgOEOSandTC} that reports phonon hardening. While we achieve agreement on theory side for CeO$_2$, no data on optical phonon frequency shift is found in literature for CeO$_2$. However, a very recent experimental study on MgO~\cite{Calandrini2021MgOOpticalModes} reports phonon softening and supports our prediction on TO phonon frequency shift, as shown in Fig.~\ref{fig.phonon}(c). We observe that the calculated frequency shift agrees better with measured value at low temperatures, but is larger than the measured value at 1500~K.

\subsection{Phonon scattering rates and linewidth}

\begin{figure*}[ht]
\centerline{\includegraphics[width=7.0in]{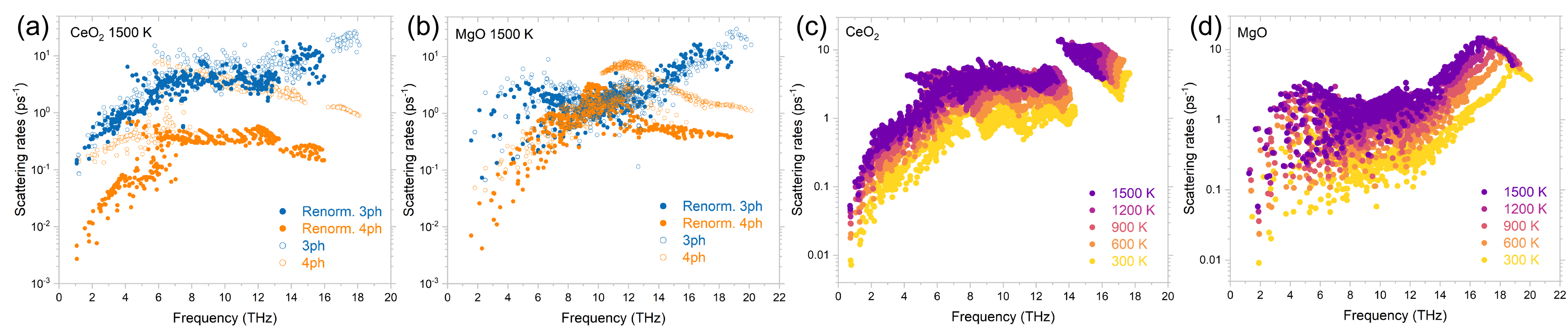}}
\caption{Phonon scattering rates of (a,c) CeO$_2$ and (b,d) MgO as a function of phonon frequency. 
Panels a-b show phonon scattering rates at 1500~K including contributions from 3ph (blue circles) and 4ph (orange circles) scattering with and without phonon renormalization. Hollow circles are scattering rates calculated without phonon renormalization (that is, with $\Phi_2$, $\Phi_3$, and $\Phi_4$), while filled circles include phonon renormalization (from the temperature-dependent $\Phi_2^{*}$, $\Phi_3^{*}$ and $\Phi_4^{*}$). 
Panels c-d show the temperature evolution of phonon scattering rates considering both 4ph scattering and phonon renormalization. Different temperatures are represented by colors with darker color indicating higher temperature.}
\label{fig.phononscattering}
\end{figure*}

Next, we move to the evaluation of phonon anharmonicity. The renormalized phonon energy shows the temperature-dependence of the real part of phonon self-energy, the scattering rates reveal its change in the imaginary part. In this study, we calculate both 3ph and 4ph scattering rates with the temperature-dependent $\Phi_2^{*}$, $\Phi_3^{*}$, and $\Phi_4^{*}$. The results for both materials are summarized in Fig.~\ref{fig.phononscattering}. Figure~\ref{fig.phononscattering}(a) and (b) present 3ph and 4ph scattering rates at 1500~K and how these two channels are affected by phonon renormalization. The calculations with $\Phi_2$, $\Phi_3$, and $\Phi_4$ (hollow circles) give high 4ph scattering rates that can be comparable or even overweigh 3ph scattering rates. The 4ph scattering strength scales quadratically with temperature~\cite{4ph2017} while 3ph scattering has a linear scaling relation, so that 4ph scattering generally becomes more important at higher temperature. However, our phonon renormalization approach with $\Phi_2^{*}$, $\Phi_3^{*}$, and $\Phi_4^{*}$ weakens 4ph scattering rates in both materials (see the difference between filled and hollow orange circles). The results for CeO$_2$ differ from MgO as renormalization makes 4ph scattering at least one magnitude lower than 3ph in CeO$_2$, while MgO still sees comparable 4ph scattering rates in intermediate phonon frequency range. This observation suggests that effect from phonon renormalization is material dependent. Another observation is that 3ph and 4ph scattering show different responses to temperature. We see greater reduction in 4ph scattering than 3ph scattering when the temperature modification is applied.

In Fig.~\ref{fig.phononscattering}(c) and (d), we show the phonon scattering rates at different temperatures up to 1500~K. The results are from temperature-dependent $\Phi_2^{*}$, $\Phi_3^{*}$, and $\Phi_4^{*}$, where both 4ph scattering and phonon renormalization effects are included. For both materials, the general trend is increasing scattering with temperature, while phonon spectrum shifts to lower frequencies (leftward) due to phonon softening at higher temperature.

While the whole spectrum phonon lifetime is not well accessible by current experimental techniques, a few optical phonon modes can be readily probed by Raman or Infrared (IR) spectroscopy providing some verification of the predicted trends. The linewidths of resonance peaks are caused by the anharmonic scattering of Raman/IR-active phonon modes, which are usually at zone-center, or phonon linewidth $\gamma = \frac{2\pi}{\tau}$~\cite{scattering1983,raman1984,4phlinewidth2020}. Comparing our simulated phonon linewidths to available measurements provides a good validation of our theory.

\begin{figure}[h]
    \centerline{\includegraphics[width=3.5in]{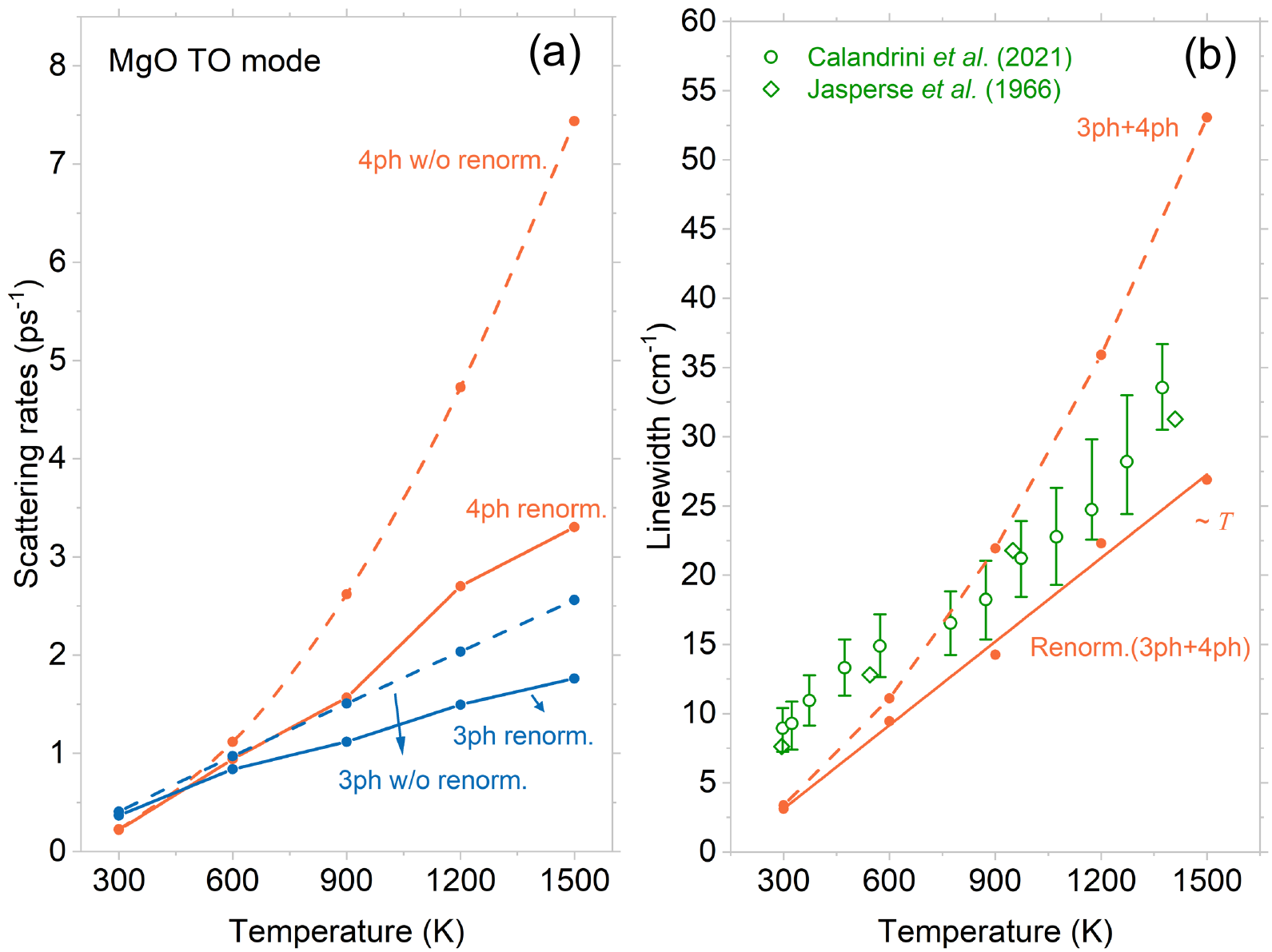}}
    \caption{Transverse optical phonon (a) scattering rates and (b) linewidth of MgO as a function of temperature. Scattering rates are calculated considering different phonon scattering mechanism (3ph in blue circles and 4ph in orange circles). Data calculated \textit{without} phonon renormalization (w/o renorm.) are connected by dashed lines, while those \textit{with} phonon renormalization (renorm.) are connected by solid lines. Total phonon linewidths are compared to experiments (green circles~\cite{Calandrini2021MgOOpticalModes} and diamonds~\cite{Jasperse1966InfraredMgO}). Data calculated \textit{without} phonon renormalization are connected by dashed lines, while those \textit{with} phonon renormalization are connected by a linear fitting (solid line).}
    \label{fig.phononlindwidth}
\end{figure}

Figure~\ref{fig.phononlindwidth} shows the TO phonon linewidth in MgO as a function of temperature. This mode is IR-active and can be detected by infrared reflectivity measurement~\cite{Calandrini2021MgOOpticalModes,Jasperse1966InfraredMgO}. We first compare the calculated scattering rates from 3ph and 4ph scattering channels and with or without the effect of phonon renormalization (Fig.~\ref{fig.phononlindwidth}(a)). Inspecting the difference between dashed lines and solid lines, the renormalization scheme reduces the scattering rates for both 3ph and 4ph scattering channels and such reduction is greater at higher temperature. Thus, this renormalization scheme is quite necessary for high-temperature understanding. Inspecting the temperature dependence of phonon scattering, 4ph scattering is still more important at higher temperature and its contribution to linewidth starts to dominate at around 600~K for MgO. However, the previously assumed quadratic temperature dependence of 4ph scattering~\cite{4ph2017} is softened to be almost linear. The temperature scaling of 3ph scattering remains to be linear after renormalization. This results in a linear temperature dependence of total phonon linewidth up to 1500~K as shown in Fig~\ref{fig.phononlindwidth}(b), and our prediction incorporating both 4ph scattering and phonon renormalization effect agrees well with a recent measurement~\cite{Calandrini2021MgOOpticalModes} (presented in green open circles) and one historic literature data~\cite{Jasperse1966InfraredMgO} (presented in green diamonds). The difference between our prediction and the experiments does not seem to be temperature-dependent and could be attributed to finite resolution in the measurements. In contrast, without renormalization 3ph+4ph scattering with $\Phi_2$, $\Phi_3$, and $\Phi_4$ gives an overestimation of phonon linewidth (dashed line in Fig~\ref{fig.phononlindwidth}(b)) while 3ph scattering with $\Phi_2$, $\Phi_3$ underestimates the linewidth as also discussed in Ref.~\cite{Calandrini2021MgOOpticalModes}. Thus, only the theoretical consideration of 4ph scattering with phonon renormalization gives a reasonable agreement with the experiments.

\begin{figure}
\centering{\includegraphics[width=3.8in]{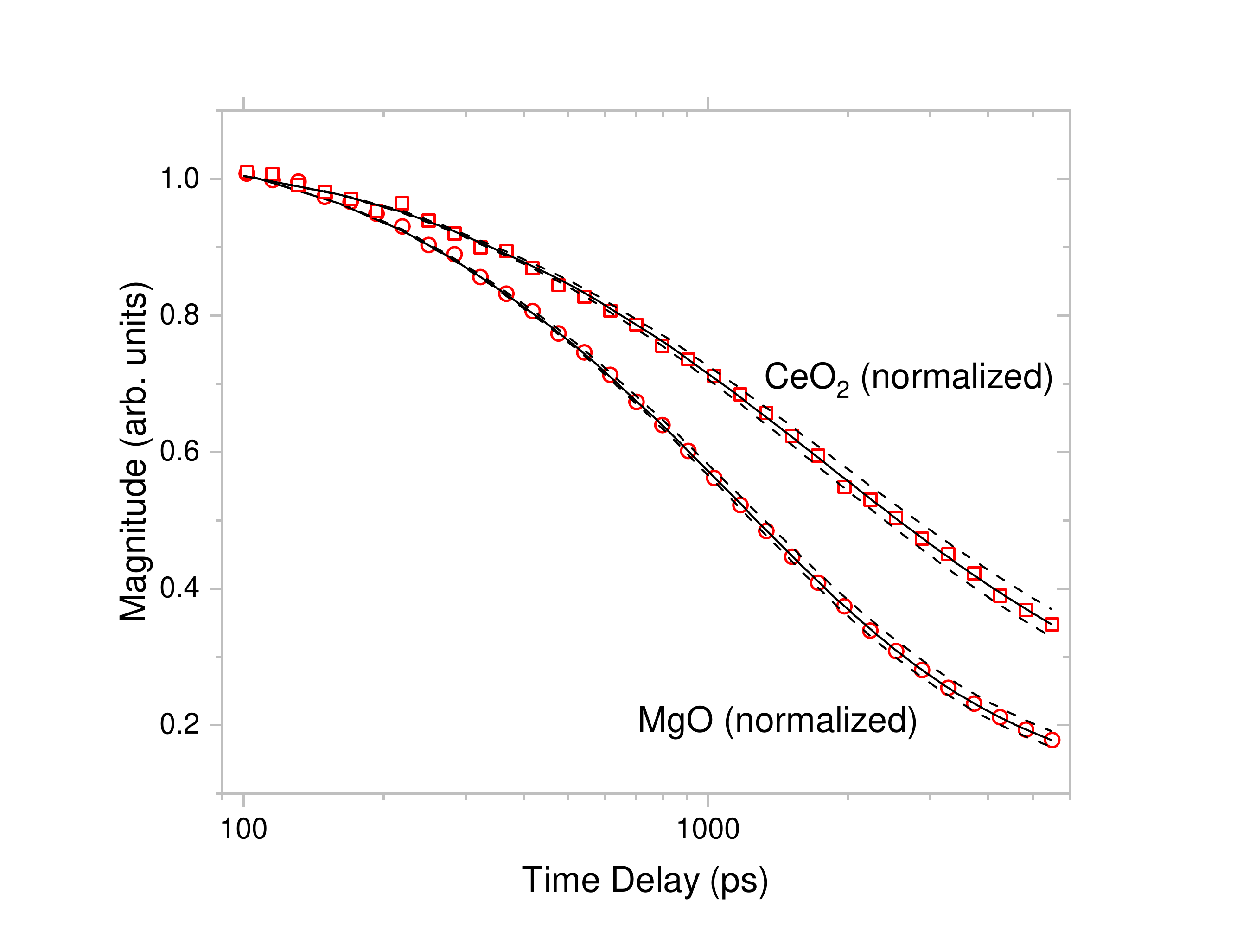}}
\caption{Normalized magnitude of thermal response as a function of time delay for TDTR measurements of MgO (red circles) and CeO$_2$ (red squares), and the numerical fits (black solid lines) used to extract their thermal conductivity. The dashed lines correspond to $\pm15\%$ variation in the fitted thermal conductivity. Note that the time delay is plotted on a logarithmic scale.}
\label{fig.TDTRfit}
\end{figure}

\begin{figure*}[!htb]
\centerline{\includegraphics[width=6.8in]{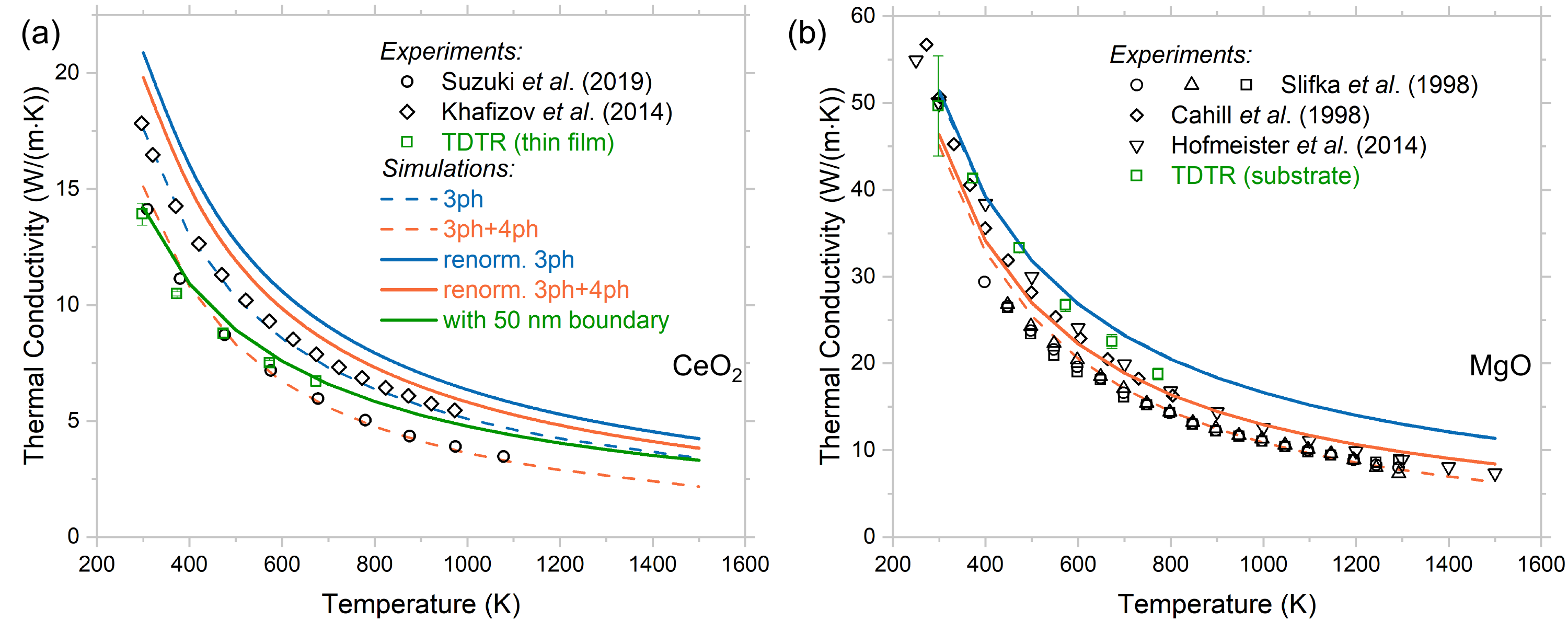}}
\caption{Thermal conductivity $\kappa$ as a function of temperature for (a) CeO$_2$ and (b) MgO. 
In panel (a), experimental data from our TDTR measurements of a thin film of CeO$_2$ (green hollow squares) are compared to literature values (from Ref.~\cite{Suzuki2019CeO2kappa} (open circles) and Ref.~\cite{Khafizov2014CeO2kappa} (open diamonds)). 
In panel (b), experimental data for MgO in literature (from Refs.~\cite{Slifka1998MgOkappa,Hofmeister2014MgOkappa,Cahill1998MgOkappa} shown in black hollow symbols) are compared to our own TDTR measurements (green squares). 
For the simulation results, dashed lines show results \textit{without} phonon renormalization and solid lines show results \textit{with} phonon renormalization (3ph in blue and 3ph+4ph in orange). 
The solid green line in (a) represents the prediction including boundary scattering (assuming 50~nm characteristic length scale). These simulation results are fitted into power law with temperature $\sim T^{-m}$.}
\label{fig.TC}
\end{figure*}

\subsection{Thermal conductivity}

The lattice thermal conductivity $\kappa$ is solved by the linearized PBTE~\cite{shengbte,han2021fourphonon} and in this study, 3ph scattering is iterated while 4ph is treated at the RTA level. We compare our simulated results to our measurements using TDTR for a CeO$_2$ thin film and an MgO substrate at high temperature, as well as to data from literature.

For our TDTR measurements, examples of the fitted model to the measured data are shown in Fig.~\ref{fig.TDTRfit}. The fitted models with $\pm$15$\%$ perturbation on fitted thermal conductivity are plotted as dashed lines to illustrate sensitivity. To address the spot-to-spot variability and effects due to varying experimental setup conditions from day to day, we report the standard deviation of the best-fit values across measurements taken at various locations on different days as error bars in Fig~\ref{fig.TC}. Noise in the measured data is negligible as the measurements are clean. 

Figure~\ref{fig.TC} summarizes our calculated results without and with renormalization (dashed and solid lines, respectively) compared with literature data (black hollow symbols) and our own TDTR measurements (green squares). In both cases we see a reduction in thermal conductivity when including the 4ph effects and the temperature dependence is further modified from $\sim T^{-1.2}$ to $\sim T^{-1}$ after phonon renormalization.

For CeO$_2$, the predicted thermal conductivity increases when renormalization is considered for both 3ph and 3ph+4ph calculations. This is consistent with the observations in Fig.~\ref{fig.phononscattering}(a) where increasing temperature reduces 4ph scattering strength compared to the case without renormalization. Our results also align with the observations of prior studies~\cite{xia2018revisiting,ravichandran2018unified,Yang2022ReducedAnharmonic} which show reduced scattering and increased thermal conductivity after renormalization.

Note that for CeO$_2$, the 3ph data (the scheme with $\Phi_2$, $\Phi_3$, or $\Phi_4$ (dashed lines)) gives accidental agreements with some literature data~\cite{Khafizov2014CeO2kappa,Suzuki2019CeO2kappa}, but that model does not provide a reasonable physical picture when we consider real experimental factors as discussed below. Notably, the thermal conductivity of CeO$_2$ is strongly dependent on grain size and purity (both hard to control in material fabrication process), while our simulations focus on single crystal, pure materials. 
Khafizov \textit{et al.}~\cite{Khafizov2014CeO2kappa} observed a reduction in thermal conductivity from 17.5~W/(m$\cdot$K) for a pellet with 5~$\mu$m average grain diameter to 7.3~W/(m$\cdot$K) when grain size decreases to 0.5~$\mu$m.  Similarly, Suzuki \textit{et al.}~\cite{Suzuki2019CeO2kappa} found that 1\% difference in purity level results in more than 50\% reduction in $\kappa$ (from 14~W/(m$\cdot$K) to 6.1~W/(m$\cdot$K)). Note that only the highest thermal conductivity results for each author are shown in Fig.~\ref{fig.TC}(a) since our simulations are for pristine materials. We also evaluated a commercially available CeO$_2$ substrate samples from PI-KEM (see Supplemental Materials for more details~\cite{supply}) observing a low thermal conductivity corresponding to a low purity level of the sample. Thus, due to this uncertainty, we focus on our high quality 800~nm grown on STO substrate by pulsed laser deposition (PLD) method to compare with the predicted results.

To further validate our temperature-dependent prediction, we estimate the average grain size of our thin film sample from XRD data to be between 40~nm and 65~nm. Thus, we compute a boundary scattering term~\cite{Mingo2005CNT}:
\begin{equation}
    \Gamma_b=\frac{2|v_\lambda^x|}{L},
\end{equation}
with characteristic length $L=50$~nm and group velocity at the transport direction $v_\lambda^x$. The boundary scattering term $\Gamma_b$ is added to the mode scattering rate $\tau_\lambda^{-1}$ in our solver and iterated with other scattering channels. The updated predictions with this boundary scattering are marked as green filled squares in Fig.~\ref{fig.TC}(a) and agree very well with our measurements on thin film samples (shown in green hollow squares).

For MgO (Fig.~\ref{fig.TC}(b)), we find good consistency in reported measurements and our experiment on MgO substrate purchased from MTI Inc. Similar to CeO$_2$, including 4ph effects reduces thermal conductivity. But for MgO, including renormalization yields a small change in the thermal conductivity. 
Overall, the prediction with renormalized 3ph and 4ph scattering gives the best agreement with the experiments, especially at higher temperatures. From 1200~K and above, scheme with $\Phi_2$, $\Phi_3$ and $\Phi_4$ (dashed orange line) gives slightly lower $\kappa$ than experiments while theoretical prediction with $\Phi_2^{*}$, $\Phi_3^{*}$ and $\Phi_4^{*}$ well captures the high temperature $\kappa$.

\section{Conclusion}

In this study, we calculated thermal properties of MgO and CeO$_2$ at temperature up to 1500 K by combining AIMD with the PBTE. The key to this scheme is computing the temperature-dependent force constants ($\Phi_2^{*}$, $\Phi_3^{*}$, and $\Phi_4^{*}$, which are obtained from TDEP method in this study). The calculated thermal properties are compared with experiments in literature and our own measurements. Both materials have positive thermal expansion with almost linear behavior for CeO$_2$ and temperature dependent behavior for MgO due to more pronounced quantum effect in MgO. The phonon frequencies soften for both materials with increasing temperature. We calculate both 3ph and 4ph scattering rates with temperature modification. Although 4ph scattering rates can be comparable to 3ph scattering rates at high temperature, our phonon renormalization approach weakens both 3ph and 4ph scattering strength. As a result, the phonon linewidths can have linear temperature dependence and our prediction explains the published data for an IR measurement on MgO. When both temperature dependence and renormalization are considered, the scattering rates of both materials increase with temperature and phonon spectrum shifts to lower frequencies due to phonon softening at high temperatures. 

The predicted decreasing trend of thermal conductivity with temperature agrees well with experiments in literature and our TDTR measurements. Due to the strong dependence on purity and grain size and other effects such as oxygen potential and stoichiometry, the experimentally determined thermal conductivity of CeO$_2$ is lower than our theoretical prediction. For CeO$_2$, the results agree well with our TDTR measurements of a thin film sample When adding a 50~nm boundary scattering, comparable to the measured grain size, to the model including $\Phi_2^{*}$, $\Phi_3^{*}$, and $\Phi_4^{*}$. Consistency between experimental measurements and prediction for bulk, single crystal MgO with the best agreement at high temperature from the model including renormalized 3ph and 4ph scattering. 

To conclude, through rigorous first-principles modeling and comprehensive comparisons with experiments on various thermal transport properties, we demonstrate the applicability of our methodology on high temperature thermal transport in ceramics. In addition, this temperature-dependent approach can reveal the temperature evolution of thermal expansion, frequency shift, and linewidth broadening, which are not accessible or incorrectly described by conventional theory.

\begin{acknowledgments}
Z.H., Z.X., J.S., H.W., A.M. and X.R. acknowledge the support from the Defense Advanced Research Projects Agency under contract number HR00112190006. The views, opinions and/or findings expressed are those of the author and should not be interpreted as representing the official views or policies of the Department of Defense or the U.S. Government. M.S. and X.X. acknowledge the support from the National Science Foundation (CBET-2051525). W.R., H.S. and P.H. appreciate support from the Office of Naval Research, Grant Number N00014-21-1-2477.

X.R. and Z.H. thank Wenjiang Zhou and Prof. Te-Huan Liu for helpful discussions on temperature-dependent force constants with polar corrections. Simulations were performed at the Rosen Center for Advanced Computing (RCAC) of Purdue University. 
\end{acknowledgments}


\bibliography{Reference}

\end{document}


\title{Supplemental Material for ``Predictions and Measurements of Thermal Conductivity of Ceramic Materials at High Temperature"}
\author{Zherui Han}
 \altaffiliation{These authors contributed equally to this work.}
 \affiliation{School of Mechanical Engineering and the Birck Nanotechnology Center,\\
Purdue University, West Lafayette, Indiana 47907-2088, USA}

\author{Zixin Xiong}
 \altaffiliation{These authors contributed equally to this work.}
 \affiliation{School of Mechanical Engineering and the Birck Nanotechnology Center,\\
Purdue University, West Lafayette, Indiana 47907-2088, USA}

\author{William T. Riffe}
\affiliation{Department of Materials Science and Engineering, University of Virginia, Charlottesville, Virginia 22904, USA}

\author{Hunter B. Schonfeld}
\affiliation{Department of Mechanical and Aerospace Engineering, University of Virginia, Charlottesville, Virginia 22904, USA}
 
\author{Mauricio Segovia}
\affiliation{School of Mechanical Engineering and the Birck Nanotechnology Center,\\
Purdue University, West Lafayette, Indiana 47907-2088, USA}

\author{Jiawei Song}
\affiliation{School of Materials Engineering, Purdue University, West Lafayette, Indiana 47907-2088, USA}
 
\author{Haiyan Wang}
 \affiliation{School of Materials Engineering,
Purdue University, West Lafayette, Indiana 47907-2088, USA}

 \author{Xianfan Xu}
 \affiliation{School of Mechanical Engineering and the Birck Nanotechnology Center,\\
Purdue University, West Lafayette, Indiana 47907-2088, USA}

\author{Patrick E. Hopkins}
 \affiliation{Department of Mechanical and Aerospace Engineering, University of Virginia, Charlottesville, Virginia 22904, USA}
 \affiliation{Department of Materials Science and Engineering, University of Virginia, Charlottesville, Virginia 22904, USA}
 \affiliation{Department of Physics, University of Virginia, Charlottesville, Virginia 22904, USA}

\author{Amy Marconnet}
 \email{marconnet@purdue.edu}
 \affiliation{School of Mechanical Engineering and the Birck Nanotechnology Center,\\
Purdue University, West Lafayette, Indiana 47907-2088, USA}

\author{Xiulin Ruan}%
 \email{ruan@purdue.edu}
 \affiliation{School of Mechanical Engineering and the Birck Nanotechnology Center,\\
Purdue University, West Lafayette, Indiana 47907-2088, USA}

\date{\today}

\maketitle
\clearpage
\tableofcontents

\section{First-principles calculations}

All calculations are done using Density Functional Theory (DFT), Density Functional Perturbation Theory (DFPT) or \textit{ab initio} molecular dynamics (AIMD) as implemented in the \textsc{VASP} package~\cite{VASP1993}. The first-principles settings for two ceramics are summarized in the table below:

\begin{table}[h]
\begin{tabular}{c|c|c|c}
\hline
Materials & XC functional & Optimization $k$-grid& Convergence criterion / (eV/\r{A}) \\ \hline
CeO$_2$ & LDA+U, $U_{\rm eff}=10~\rm eV$~\cite{HubbardU1991,LDA+UPRB2007}&   MK $12\times12\times12$ &  $10^{-6}$  \\ \hline
MgO & PBE~\cite{PBE1996}  &   $\Gamma$-centered $11\times11\times11$  &  $10^{-7}$\\ \hline
\end{tabular}
\end{table}

Energy cutoffs for both materials are 520~eV. Born effective charges are computed by DFPT and with the aid of Phonopy~\cite{phonopy} we get (1) for CeO$_2$: $\epsilon_{\infty}=5.927$, $Z^*_{Ce,xx}=5.502$ and $Z^*_{O,xx}=-2.751$; (2) for MgO: $\epsilon_{\infty}=3.241$, $Z^*_{Mg,xx}=1.980$ and $Z^*_{O,xx}=-1.980$.

\textit{Ab initio} molecular dynamics are performed on a supercell structure consisting of 192 atoms constructed by $4\times4\times4$ CeO$_2$ primitive cells or 128 atoms of $4\times4\times4$ MgO primitive cells. Only the $\Gamma$ point is computed to accelerate the calculation. After reaching thermal equilibrium under NPT ensemble (zero external pressure) with Langevin thermostat, we use 1000 more steps to get averaged lattice structure at each temperature. Then, on relaxed structure we perform NVT ensemble simulations and after reaching equilibrium we use 2000 more steps to construct effective force constants~\cite{TDEP2013IFCs} at a time step of 2~fs. In the evaluation of force constants, cutoff radii are summarized in the table below:

\begin{table}[h]
\begin{tabular}{c|c|c|c}
\hline
Materials & $\Phi_2^*$ cutoff / \r{A} & $\Phi_3^*$ cutoff / \r{A} & $\Phi_4^*$ cutoff / \r{A} \\ \hline
CeO$_2$   & 6.31~\r{A} &   6~\r{A} &  4~\r{A} \\ \hline
MgO & 4.81~\r{A}  &   4.5~\r{A}  &  3.2~\r{A} \\ \hline
\end{tabular}
\end{table}

\section{TDTR sensitivity of MgO and CeO2 on STO samples}
\begin{figure}[!htb]
    \centering
    \includegraphics[width=6.8in]{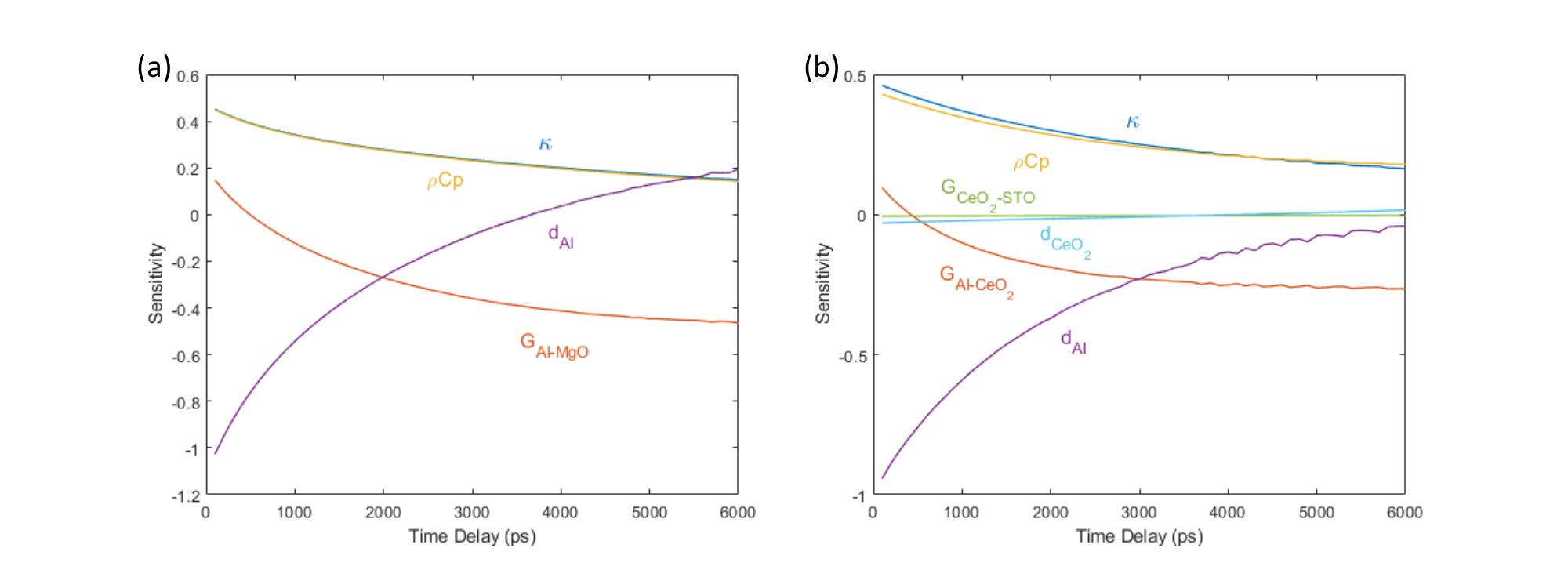}
    \caption{The sensitivity of TDTR signal to selected parameters in (a)MgO substrate and (b)CeO$_2$ film on STO substrate samples. A larger absolute value of the sensitivity corresponds to higher sensitivity to a parameter. Parameter G$_{X-Y}$ is the interface thermal conductance between material X and Y. Parameter d$_X$ is the thickness of the material X.}
    \label{fig:SI_sensitivity}
\end{figure}

The sensitivity of the TDTR signal to a parameter $\alpha$ is calculated using the expression:
\begin{equation}
    S_\alpha=\frac{d\ln{\sqrt{V_{in}^2}+V_{out}^2}}{d\ln{\alpha}}.
\end{equation}
Sensitivities to selected parameters for MgO substrate and CeO$_2$ film on STO substrate are plotted in Fig.~\ref{fig:SI_sensitivity}. The sensitivity value at zero indicates the signal is independent of the parameter. On the other hand, the signal strongly depends on parameters with large sensitivity values. For both samples, the thermal conductivity of the material of interest has high sensitivity. Therefore, the fitted thermal conductivities of MgO and CeO$_2$ have small uncertainties.

\section{TDTR measurements of CeO$_2$ substrate}
\begin{figure}[!htb]
    \centering
    \includegraphics[width=5.8in]{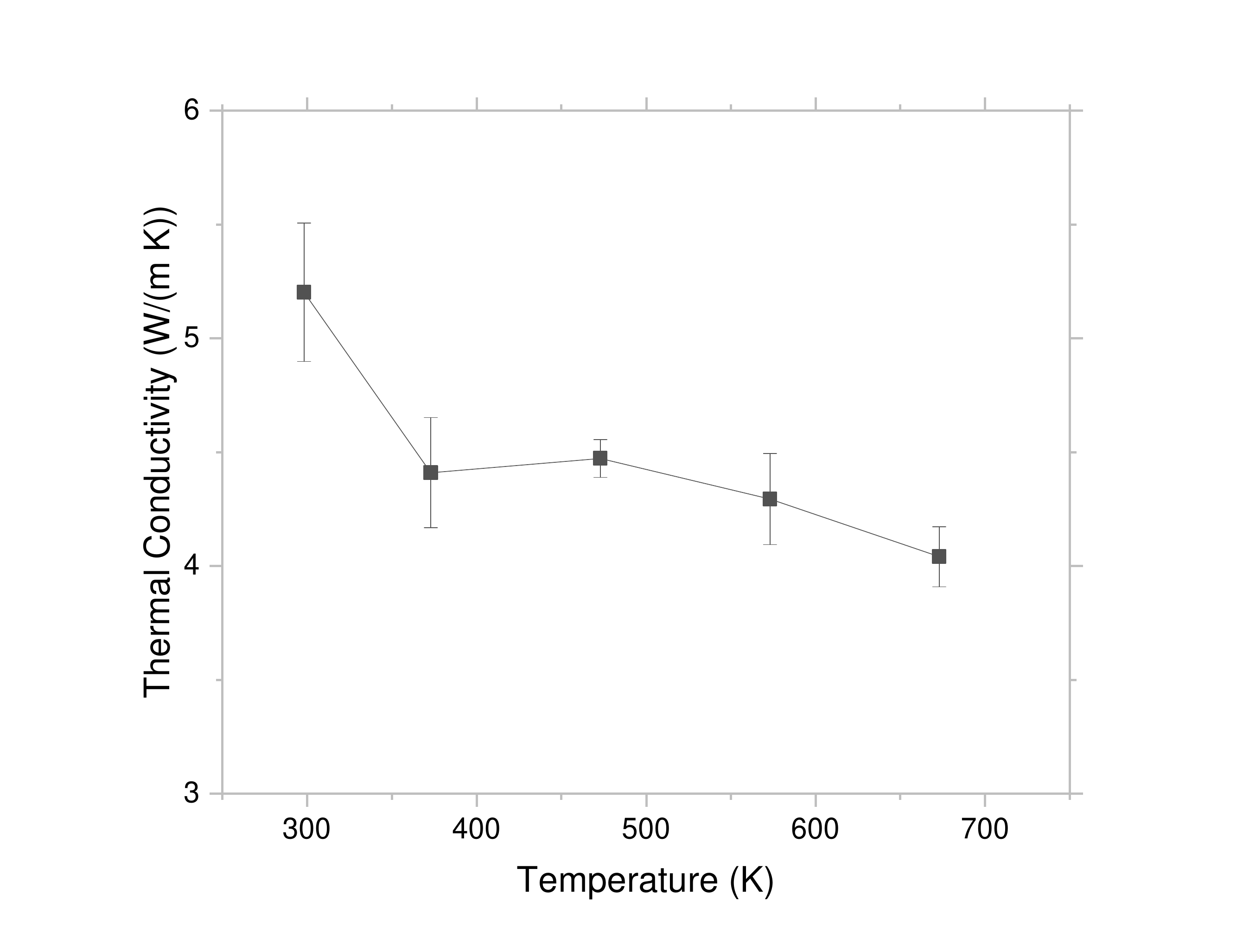}
    \caption{The measured temperature-dependent thermal conductivity of a commercially available CeO$_2$ substrate. Low thermal conductivity is likely caused by low purity, which is estimated at $~98\%$ by comparing with data from literature\cite{Nelson2014Ceo2stochiometric}}.
    \label{fig.CeO2substrate}.
\end{figure}

Due to its low purity level, the purchased CeO$_2$ substrate shows low thermal conductivity as measured by TDTR method (Fig.~\ref{fig.CeO2substrate}). The measured temperature-dependent thermal conductivity is comparable with that of CeO$_2$ with 98$\%$ purity~\cite{Nelson2014Ceo2stochiometric}.

\section{Grain size estimation based on XRD}
\begin{figure}[!htb]
    \centering
    \includegraphics[width=6.8in]{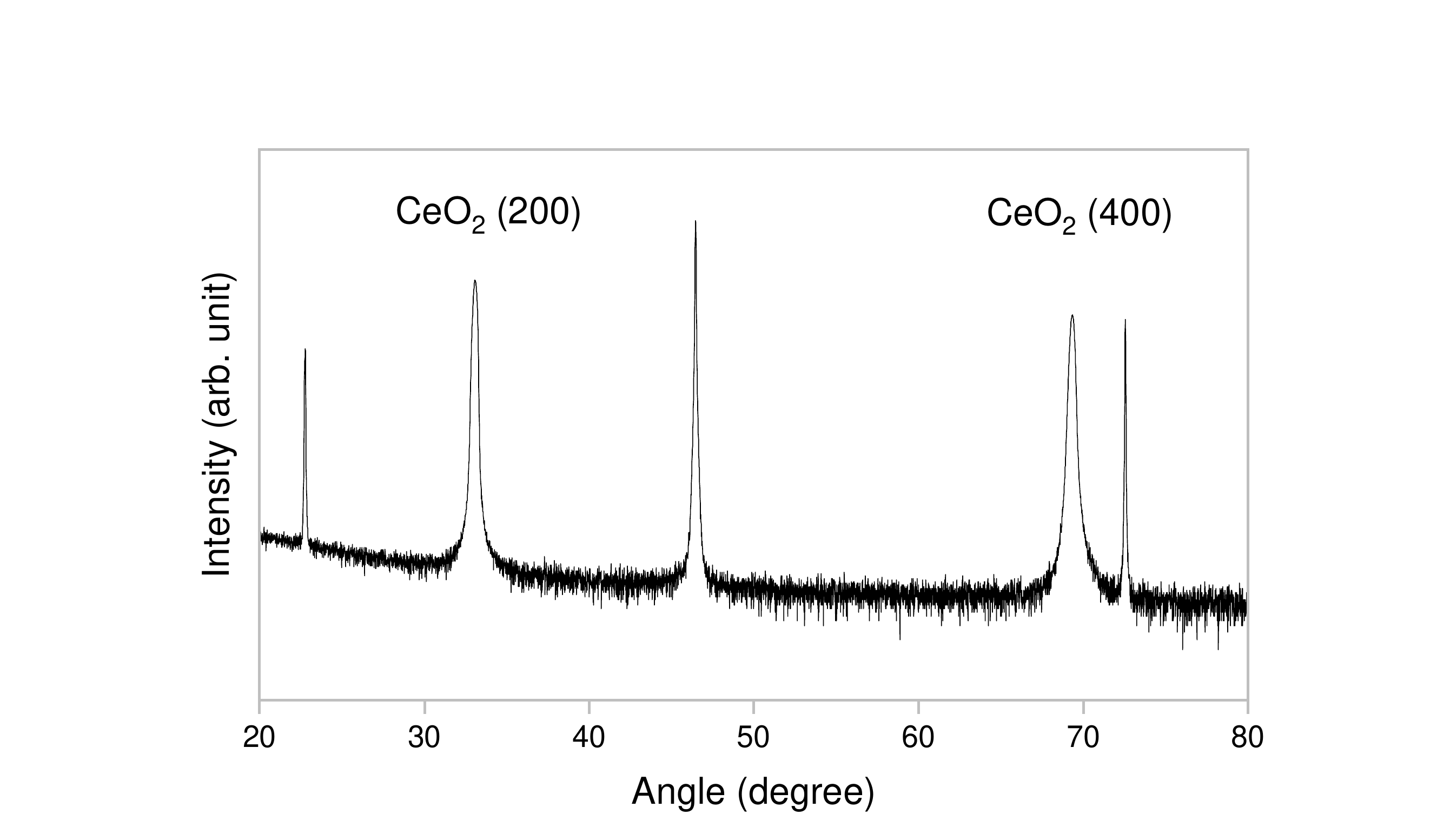}
    \caption{The XRD result of PLD-grown CeO$_2$ film on STO substrate.}
    \label{fig.XRD}
\end{figure}

To characterize the crystallinity of the CeO$_2$ substrate, XRD (Panalytical X’Pert X-ray Diffractometer with Cu K$\alpha$1 ($\lambda$=0.154 nm) radiation source) was conducted. The crystallite size of the PLD-grown CeO$_2$ film is estimated using the Scherrer Equation:
\begin{equation}
    d=\frac{k\lambda}{\beta cos\theta},
\end{equation}
where d is the average crystallite size, k is a constant with variable magnitude depending on geometry of the crystallite, $\lambda$ is the wavelength of the X-ray, $\beta$ is the full-width half-maxima of the XRD signal peak, and $\theta$ is the diffraction angle of the where the peak exists. Figure.~\ref{fig.XRD} shows the XRD result of CeO$_2$ film on STO substrate. The estimated crystallite size is between 40~nm to 65~nm, based on which the boundary scattering is calculated.

\section{Sample preparations and characterizations}
The CeO$_2$ thin film was grown on STO (001) substrate, using a pulsed laser deposition (PLD) technique with a KrF excimer laser (Lambda Physik, $\lambda$ = 248~nm, 10Hz). Before deposition, the chamber was pumped to vacuum ($<$1 × 10$^{-6}$~Torr), and substrate temperature was kept at 600 $^{\circ}$C.  A 20~mTorr oxygen pressure was used during deposition and the chamber was naturally cooled down to room temperature at 20~mTorr oxygen partial pressure after deposition.

\bibliography{Reference}